\documentclass[12pt]{iopart}

\usepackage{graphicx}
\usepackage{comment}
\usepackage{caption}
\begin{document}

\title[Measurement uncertainties and non-local quantum correlations]{Experimental investigation of the relation between measurement uncertainties and non-local quantum correlations}

\author{Kengo Matsuyama, Holger F. Hofmann and Masataka Iinuma}

\address{Graduate School of Advanced Science and Engineering, Hiroshima University, 
1-3-1 Kagamiyama, Higashi-Hiroshima, 739-8530, Japan}

\ead{matsuyama@huhep.org}

\begin{abstract} 
  Bell's inequalities are defined by sums of correlations involving non-commuting observables in each of the two systems. Violations of Bell's inequalities are only possible because the precision of any joint measurement of these observables will be limited by quantum mechanical uncertainty relations. In this paper we explore the relation between the local measurement uncertainties and the magnitude of the correlations by preparing polarization entangled photon pairs and performing joint measurements of non-commuting polarization components at different uncertainty trade-offs. The change in measurement visibility reveals the existence of a non-trivial balance between the measurement uncertainties where the probabilities of a specific pair of measurement outcomes approaches zero because of the particular combination of enhancement and suppression of the experimentally observed correlations. The occurrence of these high-contrast results shows that the quantum correlations between the photons are close to their maximal value, confirming  that the Cirel'son bound of Bell's inequality violations is defined by the minimal uncertainties that limit the precision of joint measurements.
\end{abstract}

\vspace{2pc}

\maketitle

\section{Introduction}

Violations of Bell's inequalities demonstrate that conventional quantum mechanics is incompatible with local realism \cite{PhysicsPhysiqueFizika.1.195}. Experimental violations of Bell's inequalities were first observed in optical experiments performed in the 1980s \cite{PhysRevLett.49.91}, followed by a number of refinements of the methods resulting in an increasingly reliable confirmation of the essential result \cite{ou1988violation,hensen2015loophole,giustina2015significant,shalm2015strong,hensen2016loophole,kofler2016requirements,rosenfeld2017event,big2018challenging,paneru2021experimental}. 
These experimental results all confirm the validity of the mathematical formalism without explaining the physical mechanism associated with the failure of local realism \cite{griffiths2020nonlocality}. It has even been pointed out that the mathematical analogy between classical fields and probability amplitudes permits a classical violation of Bell's inequalities which is unrelated to the problem of hidden variables and realism \cite{kagalwala2013bell,mclaren2015measuring,qian2015shifting,sandeau2016experimental,gonzales2018experimental,khrennikov2020quantum,pasini2020bell}. 
In the light of such developments, it appears to be useful to take a closer look at the original intention of Bell's inequalities as a criterion for the possibility of measurement independent realities of the non-commuting physical properties of an individual quantum system. Specifically, Bell's inequalities refer to the individual outcomes of measurements that cannot be performed jointly. A genuine violation of Bell's inequalities therefore necessarily requires the use of incompatible measurements to obtain outcomes that do not have a joint probability distribution \cite{de1984alternative,de1989joint,andersson2005joint,wolf2009measurements,hirsch2018quantum,cohen2020praise,temistocles2019measurement}. In other words, violations of Bell's inequalities are only possible because the outcomes of incompatible measurements cannot be described by a joint non-contextual reality \cite{cabello2010proposal,higgins2015using,hofmann2020contextuality}. The closest approximation of a joint non-contextual reality of non-commuting observables is an uncertainty limited joint measurement of the two observables. This means that there must be a quantitative relation between the maximal permitted violation of a Bell's inequality given by the Cirel'son bound \cite{cirel1980quantum} and the uncertainty limit of joint measurements. Several theoretical studies have explored this fundamental relation between quantum uncertainties and the Cirel'son bound based on uncertainty limits imposed by the Hilbert space formalism \cite{oppenheim2010uncertainty,banik2013degree}. A specific relation to local measurement uncertainties was recently derived in \cite{hofmann2019local}, where it was shown that the quantum correlations between two qubits are constrained by the positivity of outcome probabilities in local uncertainty limited measurements applied in parallel to both qubits. It may therefore be interesting to study the characteristics of the non-classical correlations responsible for a maximal violation of a Bell's inequality using the experimentally observable statistics of uncertainty limited joint measurements.

The advantage of applying uncertainty limited joint measurements is that the results can reveal details about the statistical correlations between observables that cannot be obtained from the separately performed precise measurements. However, it is a non-trivial task to distinguish between the effects of measurement uncertainties and the intrinsic statistics of the state. A possible solution of this problem is the use of weak measurements, where the uncertainties of a sequentially performed joint measurement are strongly biased in favour of the second measurement. It is then possible to reconstruct quasi-probabilities for the input state, resulting in an explanation of Bell's inequality violations by negative values in the reconstructed probability distribution \cite{higgins2015using}. Weak measurements have also been used to demonstrate the correspondence of weak values with the ``elements of reality'' of the EPR paradox \cite{calderon2020weak}. However, the highly biased distribution of measurement uncertainties means that the relation between the non-commuting observables appears in a somewhat distorted manner in the raw data. It may therefore be desirable to explore the whole range of uncertainty limited joint measurements, including both biased and unbiased uncertainty trade-offs. 

In this paper, we present an experimental study of the joint measurement statistics of the non-commuting observables in a maximally violated Bell's inequality. We use an optical setup in which polarization entangled photon pairs are generated by spontaneous parametric down-conversion. Since the input state is close to a maximally entangled state, the intrinsic correlations of the state achieve a Bell's inequality violation close to the Cirel'son bound. The joint measurements of non-commuting polarization components are realized by appropriate settings of polarization filters, where the uncertainty balance between the two polarization components is controlled by the polarization angles associated with the four measurement outcomes. By changing the balance of the uncertainties we can clarify the relation between intrinsic uncertainties of the state and the experimentally observed statistics. In particular, we find that probabilities close to zero can only be observed for very specific balances of the uncertainties. In these instances, a highly biased trade-off between the measurement uncertainties of the joint measurements minimizes the errors observed in a corresponding set of measurement outcomes, making these outcomes particularly susceptible to the non-local correlations of the input state. The detailed analysis of the uncertainty trade-off shows that stronger correlations of the same type would necessarily result in negative experimental probabilities, demonstrating that the Cirel'son bound must be satisfied to avoid paradoxical outcomes in uncertainty limited joint measurements.

The rest of the paper is organized as follows. In sec.\ref{sec:theory}, we explain the concept of uncertainty limited joint measurements for non-commuting observables of two level system and introduce its application to a measurement of Bell correlations. In sec.\ref{sec:experiment} we describe the experimental realization of the joint measurement and the setup of the entanglement source. Experimental results for various uncertainty trade-offs are presented in sec.\ref{sec:results}. Sec.\ref{sec:conclusion} summarizes the results and concludes the paper.  

\section{Joint measurements of the correlations in Bell's inequalities}
  \label{sec:theory}

\subsection{Uncertainty limited joint measurement of two non-commuting physical properties in a two-level system}
  \label{subsec:jm}

\begin{figure}[htbp]
  \centering
  \includegraphics[width=100mm]{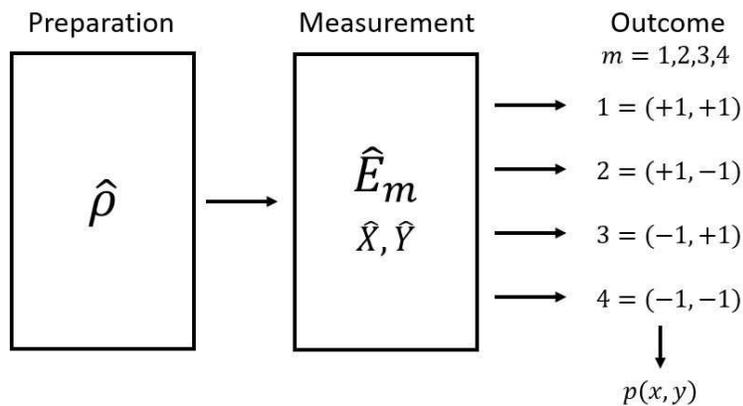}
  \caption{Illustration of a joint measurement of the two non-commuting physical properties, $\hat{X}$ and $\hat{Y}$. $\hat{\rho}$ is the initial state of the quantum system and the $\hat{E}_m $ is the element of the POVM. The indices $m$ represent the four different outcomes, $1=(+1, +1)$, $2=(+1, -1)$, $3=(-1, +1)$, and $4=(-1, -1)$. The quantum state determines the probabilities of each outcome, resulting in a joint probability distribution $p(x,y)$.}
  \label{fig:joint}
\end{figure}

Bell's inequalities typically involve pairs of non-commuting physical properties in two level systems, which are given by pairs of operators $\hat{X}$ and $\hat{Y}$ with eigenvalues of $x=\pm1$ and $y=\pm1$. It is possible to construct a joint measurement of the two non-commuting properties such that the measurement outcomes are given by the four possible combinations of eigenvalues, $(x=\pm 1, y=\pm 1)$. 
As shown in Fig. \ref{fig:joint}, the joint measurement will produce one of the four possible outcomes whenever a quantum system in a state $\hat{\rho}$ is measured. The statistics of these four measurement outcomes can be represented by a joint probability distribution, $p(x,y)$. For eigenvalues of $\pm 1$, measurement uncertainties will result in a reduction of the mean value of $x$ and $y$ with respect to the expectation values $\langle \hat{X} \rangle$ and $\langle \hat{Y} \rangle$ of the input state. This reduction factor is a convenient representation of the loss of information caused by the measurement uncertainties. In the following, these ratios will be represented by the visibilities $V_X$ and $V_Y$.
  
Joint measurements of two non-commuting properties can be designed in such a way that each measurement outcome can be represented by a combination of eigenvalues of the two observables. For two level systems, such joint measurements can be represented by a positive operator valued measure (POVM) defined by the visibilities $V_X$ and $V_Y$,
  \begin{eqnarray}
    \hat{E}_m = \frac{1}{4}(\hat{I} \pm V_{X} \hat{X} \pm V_{Y} \hat{Y}),
    \label{eqn:povm elements for joint measurement}
  \end{eqnarray}
where the measurement outcome $m$ represents one of the four possible outcomes and the signs in front of $V_{X} \hat{X}$ and $V_{Y} \hat{Y}$ correspond to the signs of the values of $x$ and $y$ for that outcome $m$. Since the operator measure must be positive, the visibilities of the measurement necessarily satisfy the uncertainty relation \cite{englert1996fringe,iinuma2011weak}
  \begin{eqnarray}
    V_{X}^2 + V_{Y}^2 \le 1. 
  \end{eqnarray}
This uncertainty relation describes a circle of radius one in the plane defined by $V_{X}$ and $V_{Y}$. It is therefore convenient to represent uncertainty limited measurements by a single parameter $\theta$, so that
  \begin{eqnarray}
    V_{X} &=& \cos \theta, \nonumber \\
    V_{Y} &=& \sin \theta. 
  \label{eqn:theta}
  \end{eqnarray}
The parameter $\theta$ thus describes the uncertainty trade-off as a balance between $\hat{X}$ and $\hat{Y}$, where $\theta=0^\circ$ is maximal $\hat{X}$-resolution and $\theta=90^\circ$ is maximal $\hat{Y}$-resolution.

\subsection{Observation of correlations using joint measurements}
\label{subsec:ocj}

In the following, we consider a pair of two-level systems, $A$ and $B$, with the local non-commuting properties $\hat{X}_A$, $\hat{Y}_A$ and $\hat{X}_B$, $\hat{Y}_B$. These properties can violate a Bell's inequality given by
\begin{eqnarray}
  |\langle \hat{X}_A \hat{X}_B \rangle - \langle \hat{X}_A \hat{Y}_B \rangle + \langle \hat{Y}_A \hat{X}_B \rangle + \langle \hat{Y}_A \hat{Y}_B \rangle | \le 2.
   \label{eqn:Bell}
\end{eqnarray}
In experimental tests of Bell's inequality violations, the four Bell correlations $\langle \hat{X}_A \hat{X}_B \rangle$, $\langle \hat{X}_A \hat{Y}_B \rangle$, $\langle \hat{Y}_A \hat{X}_B \rangle$ and $\langle \hat{Y}_A \hat{Y}_B \rangle$ all have to be evaluated separately. It is therefore impossible to observe the sum of the four Bell correlations directly. It is nevertheless possible to define a collective observable $\hat{B}$ summarizing the four correlations, so that
\begin{eqnarray}
  \hat{B} = \hat{X}_A \hat{X}_B - \hat{X}_A \hat{Y}_B  +  \hat{Y}_A \hat{X}_B  + \hat{Y}_A \hat{Y}_B. 
\end{eqnarray}
The violation of Bell's inequality is possible because the eigenstates of this operator include values of $+2 \sqrt{2}$ and $-2 \sqrt{2}$, both of which lie outside of the range of expectation values permitted by the Bell's inequality in Eq.(\ref{eqn:Bell}). In the following, we will use the state with an eigenvalue of $-2 \sqrt{2}$ to achieve a maximal violation of the Bell's inequality in Eq.(\ref{eqn:Bell}). 
  
The operator $\hat{B}$ represents a non-local quantity that cannot be directly observed in any local measurements of the two systems because it includes two pairs of non-commuting local properties, $\hat{X}_A, \hat{Y}_A$ and $\hat{X}_B, \hat{Y}_B$. However, it is possible to obtain joint information on all four local properties by performing a joint measurement of $x_A$, $y_A$ in $A$ and a joint measurement of $x_B$, $y_B$ in $B$. The outcomes of the two joint measurements define an individual value of $b$ for the correlations in Bell's inequalities. For an outcome of $(x_A,y_A;x_B,y_B)$, this $b$-value is given by
\begin{eqnarray}
  b = x_A x_B - x_A y_B  +  y_A x_B  + y_A y_B. 
\end{eqnarray}
It is easy to see that $b$ can only take values of $+2$ or $-2$ for any combination of measurement outcomes. It is therefore obvious that any joint assignment of local measurement values satisfies Bell's inequalities. The joint observation of $(x_A,y_A;x_B,y_B)$ for an input state violating Bell's inequalities is only possible because the measurement uncertainties of the joint measurements reduce the average of $b$ to a value between $-2$ and $+2$ for any possible expectation value of $\hat{B}$ in the input state. 
  
\begin{figure}[htbp]
    \centering
    \includegraphics[width=90mm]{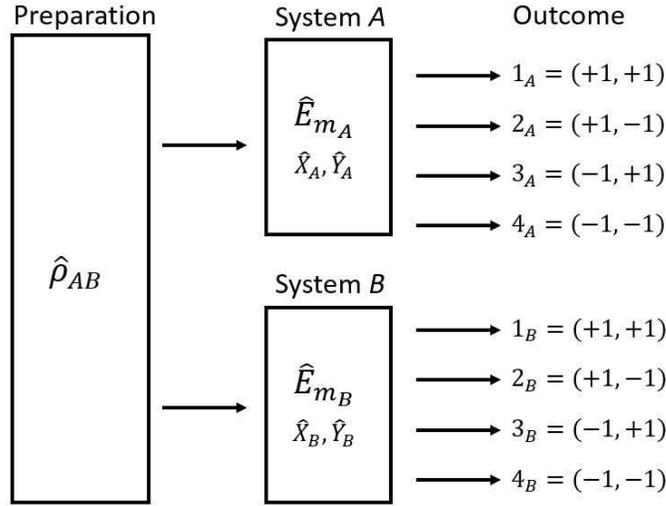}
    \caption{Illustration of a measurement of correlated pairs with joint measurements used on each of the two local systems $A$ and $B$. Each outcome is described by a combination of an outcome $m_A$ in system $A$ and $m_B$ in system $B$. There are sixteen possible outcomes, corresponding to the sixteen possible combinations of four values of $\pm 1$.}
    \label{fig:outline of the two joint measurement}
\end{figure}
  
Fig. \ref{fig:outline of the two joint measurement} shows an illustration of the joint measurement of $x_A$, $y_A$, $x_B$ and $y_B$. 
Since the two joint measurements performed in $A$ and in $B$ give four measurement outcomes each, there are a total of sixteen measurement outcomes and the statistics obtained from a specific quantum state is given by a joint probability distribution $p(x_A,y_A;x_B,y_B)$. Tab. \ref{table:B} shows the values of $b$ for each of the sixteen outcomes by cross-referencing the local outcomes in $A$ with the local outcomes in $B$. 
  
\begin{table}[htbp]
    \centering
    \caption{Table of the values of $b$ for the sixteen different combinations of local measurement outcomes in $A$ with local measurement outcomes in $B$.}
    \label{table:B}
    \footnotesize
    \begin{tabular}{@{}ccccc}
    \br
    & $x_A=+1$ & $x_A=+1$ & $x_A=+1$ & $x_A=+1$\\
    & $y_A=+1$ & $y_A=-1$ & $y_A=+1$ & $y_A=-1$\\
    \br
    $x_B=+1$ & $b=+2$ & $b=-2$ & $b=+2$ & $b=-2$\\
    $y_B=+1$ &  &  &  & \\
    \mr
    $x_B=+1$ & $b=+2$ & $b=+2$ & $b=-2$ & $b=-2$\\
    $y_B=-1$ &  &  &  & \\
    \mr
    $x_B=-1$ & $b=-2$ & $b=-2$ & $b=+2$ & $b=+2$\\
    $y_B=+1$ &  &  &  & \\
    \mr
    $x_B=-1$ & $b=-2$ & $b=+2$ & $b=-2$ & $b=+2$\\
    $y_B=-1$ &  &  &  & \\
    \br
\end{tabular}\\
    
\end{table}
\normalsize

The results of joint measurements for a given input state can be represented by a probability distribution $p(x_A,y_A;x_B,y_B)$ describing the modification of input probabilities by the measurement uncertainties. The experimentally observed Bell correlation $b$ can be evaluated by taking the average of the value $b$,
  \begin{eqnarray}
    \langle b \rangle = 2P_{\mathrm{exp.}}(b=+2)-2P_{\mathrm{exp.}}(b=-2),
    \label{eqn:average_B_value}
  \end{eqnarray}
where $P_{\mathrm{exp.}}(b=+2)$ is the sum over all probabilities of outcomes with $b=+2$ and $P_{\mathrm{exp.}}(b=-2)$ is the sum over all probabilities of outcomes with $b=-2$. Since the experimental probabilities will always be positive the result will always be within the range permitted by Bell's inequalities. The physical reason why $|\langle b \rangle|$ is lower than $|\langle \hat{B} \rangle|$ is that the measurement uncertainties reduce the expectation values of the local measurement outcomes. Effectively, measurement errors will increase the low probabilities ($P_{\mathrm{exp.}}(b=+2)$ in our case) and decrease the high probabilities ($P_{\mathrm{exp.}}(b=-2)$ in our case). The average $\langle b \rangle$ is eventually determined by the competition between the randomization caused by the measurement uncertainties and the strength of the non-local correlations represented by $\langle \hat{B} \rangle$. If the effect of measurement uncertainties was negligible, a value of $|\langle \hat{B} \rangle|>2$ would correspond to a negative probability for either $b=+2$ or $b=-2$, representing the bare statistics of an input state given by $\langle \hat{B} \rangle$. Measurement uncertainties are therefore necessary to avoid negative probabilities in the joint measurements of input states maximally violating a Bell's inequality.

\section{Experiment}
  \label{sec:experiment}
\subsection{Realization of joint measurements for two complementary polarizations}
\label{subsec:joint measurement}

The polarization of a single photon is a natural two level system, with orthogonal polarizations representing the eigenstates of a specific measurement basis. In the present experiment, we need to select polarization operators that maximally violate a Bell's inequality for a down-converted photon pair with maximal correlations between parallel polarizations. We can do this by introducing a rotation angle of $22.5^\circ$ between the polarizations of the operators for the two photons. As shown in Tab. \ref{table:physical property and polarization}, we choose horizontal ($0^\circ$) and vertical ($90^\circ$) polarizations for the eigenvalues of $+1$ and $-1$ of the operator $\hat{X}_A$, and diagonal polarizations of $45^\circ$ and $135^\circ$ for the eigenvalues of $+1$ and $-1$ of the operator $\hat{Y}_A$. A maximal violation of Bell's inequalities is then obtained by adding $+22.5^\circ$ to the orientation of each eigenstate polarization in the definitions of the operators $\hat{X}_B$ ($22.5^\circ$ and $112.5^\circ$) and $\hat{Y}_B$ ($67.5^\circ$ and $157.5^\circ$). 

\begin{table}[htbp]
  \centering
  \caption{Definition of polarization eigenstates for the operators $\hat{X}_A$, $\hat{Y}_A$, $\hat{X}_B$ and $\hat{Y}_B$. Polarizations are given by their orientation in real space, where $0^\circ$ is assigned to horizontal polarization and $90^\circ$ is assigned to vertical polarization.}
  \label{table:physical property and polarization}
  \footnotesize
  \begin{tabular}{@{}llll}
  \br
  Physical property & Value &  Angle in real space\\
  \mr
  $\hat{X}_A$ & $(+1, -1)$ &  $( 0^\circ, 90^\circ )$ \\
  $\hat{Y}_A$ & $(+1, -1)$ &  $( 45^\circ, 135^\circ )$ \\
  $\hat{X}_B$ & $(+1, -1)$ &  $(22.5^\circ, 112.5^\circ ) $\\
  $\hat{Y}_B$ & $(+1, -1)$ &  $(67.5^\circ, 157.5^\circ ) $\\
  \br
  \end{tabular}\\  
\end{table}
\normalsize

\begin{figure}[h]
  \centering
  \includegraphics[width=100mm]{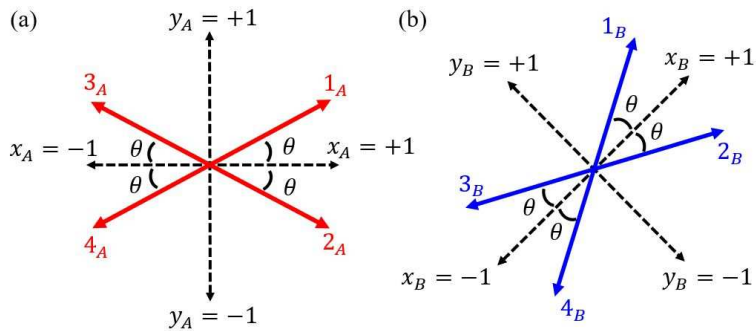}
  \caption{Bloch vector representation of the polarization directions used to implement the joint measurements of $\hat{X}$ and $\hat{Y}$. The polarization directions for photon $A$ are shown in (a), where the axes given by the operators $\hat{X}$ and $\hat{Y}$ are aligned with the horizontal and vertical directions in the graph. The polarization directions for photon $B$ are shown in (b), where the axes given by the operators $\hat{X}$ and $\hat{Y}$ are rotated by $45^\circ$ to illustrate the rotation by $22.5^\circ$ of the polarization directions in real space. The labels used for the four different measurement outcomes are identical to the ones used in Fig. \ref{fig:outline of the two joint measurement}.}
  \label{fig:povm elements}
\end{figure}

Joint measurements can be realized by selecting polarization angles between two eigenstates belonging to different operators. Each measurement must have four different outcomes, represented by combinations of eigenvalues $(x,y)$. The polarization angle is set to be the angle between the polarizations of the $\hat{X}$ eigenstates and the polarizations detected in the joint measurements and should have the same absolute value for all four measurement outcomes. The uncertainty trade-off described by Eq.(\ref{eqn:theta}) is determined by this angle between the polarization of the $\hat{X}$ eigenstate and the polarization detected in the joint measurement. The value of the angle by which the polarizations are rotated relative to the $\hat{X}$ polarization is equal to $\theta /2$ , so that all settings between $\hat{X}$ polarization and $\hat{Y}$ polarization correspond to the uncertainty trade-off described by $0^\circ \leq \theta \leq 90^\circ$. The measurement basis for the four measurement outcomes can be visualized in terms of the Bloch vector representation of the measurement operators, where $\theta$ is the angle between the axis defined by $\hat{X}$ and the Bloch vectors of the four measurement outcomes. Fig. \ref{fig:povm elements} shows the Bloch vector orientations of the four measurement outcomes, where the rotation of polarizations by $22.5^\circ$ for photon $B$ is represented by a $45^\circ$ rotation of the axes in the Bloch representation. As explained above, the actual rotation angles for the settings of the polarization filters in each measurement can be obtained by rotating the polarization direction associated with the eigenvalue of $\hat{X}$ by an angle of $\theta/2$ towards the polarization direction associated with the eigenvalue of $\hat{Y}$. In the experiment, $\theta = 0^\circ$ corresponds to a maximal visibility of both $\hat{X}_A$ and $\hat{X}_B$, and  $\theta = 90^\circ$ corresponding to a maximal visibility of both $\hat{Y}_A$ and $\hat{Y}_B$. Equal measurement visibilities for all polarizations are obtained at $\theta = 45^\circ$.

\subsection{Experimental setup}

\begin{figure}[h]
	\centering
	\includegraphics[width=100mm]{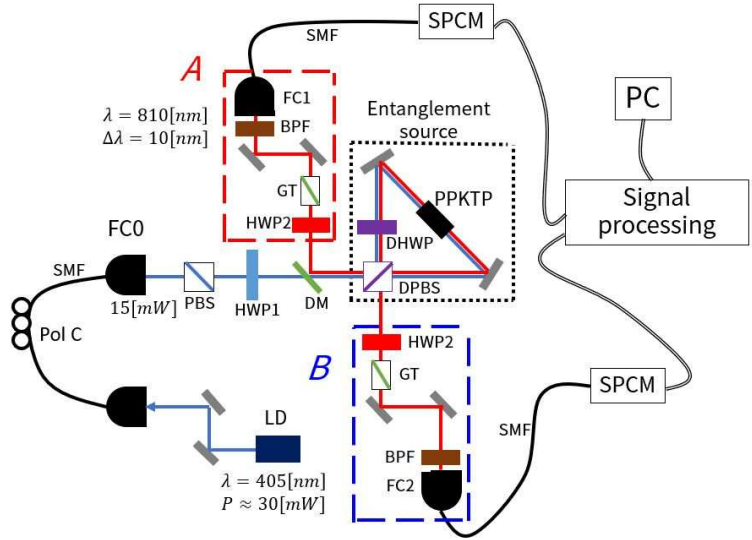}
	\caption{Experimental setup to measure the joint statistical distributions for the maximally entangled photon pairs. The pump beam is prepared using a laser diode (LD) with a wavelength of 405 nm and an output power of 30 mW. The pump beam is emitted from the fiber coupler FC0 at a power of 15 mW. The pump beam is then separated at the dual polarization beam splitter (DPBS) into two optical paths, pumping the PPKTP crystal ($1mm \times 1mm \times 10mm$) from both sides and creating photon pairs in the two paths. A double half waveplate (DHWP) is inserted to exchange horizontal and vertical polarizations with each other. The two paths then overlap at the DPBS and entangled photon pairs are emitted from the two output ports. The two photons are then detected separately by two single photon counting modules(SPCMS) after passing through polarization filters and band pass filters (BPF). The detection setup for photon $A$ is shown inside the red-dotted box and the  and the corresponding setup for $B$ is shown in the blue-dotted box. The output pulses were converted to NIM logic pulses of 30 ns and processed using a NIM logic circuit composed of Logic Level Adapter, Discriminator, and Coincident counting modules. The coincidence counts were recorded by a connected PC.}
	\label{fig:setup}
\end{figure}

The experimental setup in Fig. \ref{fig:setup} is composed of a Sagnac-type interferometer used as a source of the entangled photon pairs, two polarization filters, and the electronics used for signal processing. A pump beam with a wavelength of 405 nm is emitted from a fiber coupler (FC0) and its polarization is adjusted using a polarization beam splitter (PBS) and a half-wave plate (HWP1) to select the diagonal polarization needed for the generation of maximally polarization entangled photon pairs. The blue beam is separated at a dual polarization beam splitter (DPBS) into two optical paths circulating right and left in the Sagnac interferometer, pumping the periodically poled KTP (PPKTP) crystal from both sides and creating photon pairs in the two paths. Entanglement is generated when the down-converted two photon wave functions interfere at the DPBS as they exit the Sagnac interferometer. The entangled photon pairs exit the two output ports of the DPBS and pass through two polarization filters constructed from a HWP2 and a Glan-Taylor prism(GT). The polarization angle for each measurement setting $\theta$ is controlled by rotating the HWP2 by an angle of $\theta/4$ from the angle defined by the $\hat{X}$ eigenstate with the intended outcome eigenvalue $x$ towards the intended outcome eigenvalue $y$. There are two local measurement setups for each of the two photons, $A$(red-dotted box) and $B$(blue-dotted box). Two band pass filters (BPFs) before two fiber couplers (FC1 and FC2) eliminate background photons and select photon pairs with the same wavelength. The selected photon pairs are then detected by two single photon counting modules(SPCMS), PerkinElmer (SPCM-AQR-14-FC13237-1) and EXCELITAS (SPCM-AQRH-14-FC24360) and their electronically processed coincidence counts are recorded and stored in a computer system.

The quality of the initial state was evaluated by observing the correlations of parallel local polarizations. For a local polarization angle $\phi$, the result $+$ corresponds to a setting of $\phi$ for the polarization filter in front of the detector and the result $-$ corresponds to the orthogonal setting. The visibility $V_{\phi}$ can then be determined directly from the count rates observed in the four possible settings, 
\begin{eqnarray}
	V_{\phi} \equiv \frac{N_{+-}+N_{-+}-N_{++}-N_{--}}{N_{+-}+N_{-+}+N_{++}+N_{--}}, \nonumber
\end{eqnarray}
with $+-$ and $-+$ representing the intended anti-correlation of polarizations and $++$ and $--$ representing the errors. The two photon visibilities obtained at $ \phi = 0^\circ, 45^\circ, 90^\circ$ and $135^\circ$ were 
\begin{eqnarray}
	V_{0^\circ} = 0.980 \pm 0.004   &  \; \; \; \; \;  V_{90^\circ} = 0.969 \pm 0.004 \nonumber \\
	V_{45^\circ} = 0.976 \pm 0.004  &  \; \; \; \; \;  V_{135^\circ} = 0.975 \pm 0.004.
	\label{eqn:inter visi}
\end{eqnarray}
Note that the settings of $\phi = 0^\circ$ and $\phi = 90^\circ$ both represent the horizontal/vertical correlation and $\phi = 45^\circ$ and $\phi = 135^\circ$ both represent the diagonal polarizations. Slight deviations in the results might be an indication of the precision with which experimental settings were realized.

\subsection{Performance of joint measurement}

The performance of the joint measurements was evaluated by experimental determination of the visibilities comparing the expectation values of precise measurements with the averages obtained in joint measurements, 
\begin{eqnarray}
  V_{\xi} \equiv \frac{\langle \xi \rangle_{\mathrm{joint}}}{\langle \xi \rangle_{\mathrm{precise}}}. 
\end{eqnarray}
Here $\xi = x_A, y_A, x_B, y_B $ represents the outcomes of the measurements. $\langle \xi \rangle_{\mathrm{precise}}$ represents the average obtained in a precise measurement of the polarization performed by selecting the eigenstate polarizations for that polarization and $\langle \xi \rangle_{\mathrm{joint}}$ represents the average obtained in a joint measurement for the same input conditions. 

\begin{figure}[h]
	\centering
	\includegraphics[width = 140mm]{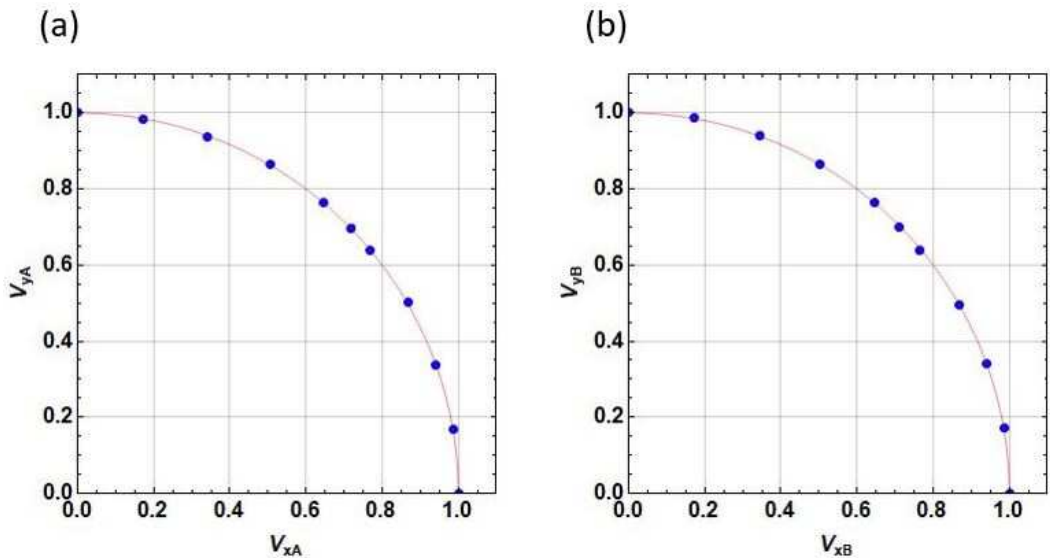}
	\caption{Experimental results for the uncertainty limit of joint measurements of $\hat{X}$ and $\hat{Y}$ in (a) system $A$ and (b) system $B$. All dots represent experimental data. Their statistical errors are sufficiently smaller than the size of the dots. The thin red curves show a circle of radius one, representing the theoretically expected uncertainty limit.}
	\label{fig:results of measurement visibilities}
\end{figure}

The most simple way of evaluating $V_{\xi}$ is to prepare an eigenstate with eigenvalue $\xi$ as the initial state and to measure it jointly. In this case, the average of the precise measurement would be given by the eigenvalue of the input state. In the present experiment, it is more convenient to use the quantum correlations of the entangled photon pairs to realize the input state. We used remote state preparation to prepare specific states in $A$ by detecting a corresponding polarization direction in $B$ and vice versa. For example, we can prepare an approximate eigenstate of $x_A = +1$ (horizontal polarization) in $A$ by detecting vertical polarization in $B$. Due to imperfections of the entanglement source this procedure prepares a mixed state and $\langle x_A \rangle_{\mathrm{precise}}$ is not equal to one. However, we can obtain precise data for its value from the settings used to evaluate the two photon interferences. 
We determined the visibilities by comparing the averages obtained from joint measurements with those obtained from precise measurements for the same conditional input state prepared by remote state preparation using the entangled photon source. 
Fig. \ref{fig:results of measurement visibilities} shows the results in two-dimensional plots that relate the visibilities $V_{X}$ and $V_{Y}$ to each other. The results for both the polarization detection in $A$ and the polarization detection in $B$ show minimal uncertainties as described by Eq.(\ref{eqn:theta}). For confirmation, we have evaluated the average distance of each point from the origin as given by $\sqrt{V_{X}^2 + V_{Y}^2}$.  The average distance was $ 1.0002 \pm 0.0017$ in $A$ and $ 1.0003 \pm 0.0015 $ in $B$, respectively.  These results indicate that we have realized a joint measurement with minimal measurement uncertainties for each of the two photons.

\section{Results and discussion}
\label{sec:results}

\subsection{Distribution observed with equal uncertainty of $\hat{X}$ and $\hat{Y}$}
\label{subsec:eu}

\begin{figure}[h]
	\centering
	\includegraphics[width=70mm]{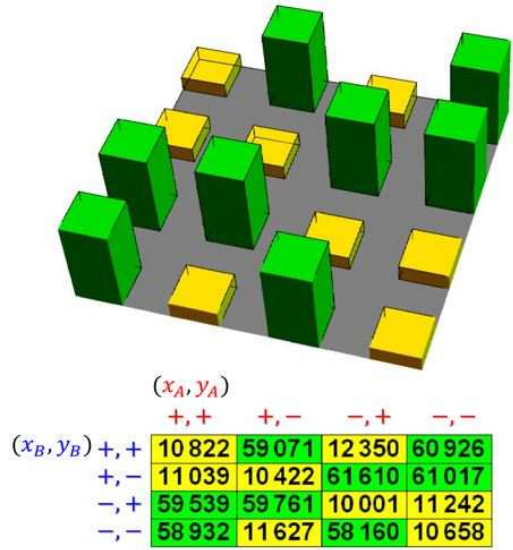}
	\caption{Raw data of the joint statistical distribution measured at $\theta = 45^\circ$, which corresponds to equal visibilities in the $\hat{X}$ and $\hat{Y}$ measurements. Outcomes with $b=+2$ are shown in yellow and outcomes with $b=-2$ are shown in green. The numbers given in the table are coincidence counts detected in 10 seconds. Outcomes are labeled in red for $A$ and in blue for $B$, with $+$ and $-$ used in place of the values $+1$ or $-1$. The bar graph is added to better visualize the distribution given in the table.}
	\label{fig:statistical distributions 45}
\end{figure}

The effects of measurement uncertainties are easiest to understand when all uncertainties are the same. In the present setup, this is the case when $\theta = 45^\circ$ for the two measurements. Fig. \ref{fig:statistical distributions 45} shows the number of counts obtained in a 10 second interval for each of the sixteen possible outcomes at $\theta = 45^\circ$. 

Since the polarization angle $\theta$ is same in both $A$ and $B$, the equal visibility makes a situation where $x_A, y_A, x_A$ and $y_B$ are measured with the same precision. To distinguish the measurement outcomes with $b=+2$ and $b=-2$, we use a color code, where yellow and green in the outcomes indicates $b = +2$ and $b = -2$ each. 

The main feature of the result at $\theta = 45^\circ$ is that the count rates of each outcome depend only on the value of $b$ associated with that outcome. As shown in Fig. \ref{fig:statistical distributions 45}, the count rates of the eight outcomes with $b = +2$ (yellow) are all equally low and the count rates of the eight outcomes with $b = -2$ (green) are all equally high. The probabilities do not depend on the precise distribution of the individual outcomes $x_A, y_A, x_B$ and $y_B$. Since the measurement errors are not biased, the preference for $b=-2$ results over $b=+2$ results can be traced back to the input state. As expected from the quantum expectation value $\langle \hat{B} \rangle=-2 \sqrt{2}$, the input state is biased in favour of negative $b$-values. 

The joint measurement at equal visibilities shows that the violation of Bell's inequalities corresponds to a bias in favour of the joint outcomes with $b$-values closest to the quantum expectation values. What is more, the state maximally violating the Bell's inequality has no other statistical biases. The intrinsic statistics of the state are completely characterized by the magnitude of the Bell's inequality violation and the associated distribution of $b$-values observed in the joint measurement. The probabilities of the $b$-values observed in the experiment are  
\begin{eqnarray}
	P_{\mathrm{exp.}}(b = +2) = 0.1554 \pm  0.0005  
  \label{eqn:b=+2 prob at 45} \\
	P_{\mathrm{exp.}}(b = -2) = 0.8446 \pm  0.0013.
	\label{eqn:b=-2 prob at 45}
\end{eqnarray}
These results correspond to an average value of $\langle b \rangle = - 1.3784 \pm 0.0028$, where the difference between $b=-2$ and the average is caused by the non-zero probability of outcomes with $b=+2$. 

It is interesting to note that the measurement uncertainties result in a significant positive probability of $b=+2$. It seems that a much larger value of $\langle \hat{B} \rangle$ might be possible without causing any problems in the joint measurement at equal visibilities. To understand why this is the case we need to consider the general effects of measurement uncertainties on the experimentally observed distributions.

\subsection{Effects of measurement uncertainties on the experimentally observed distribution}
\label{subsec:emu}

From Fig. \ref{fig:statistical distributions 45}, we can conclude that the intrinsic probability distribution of the input state is biased in favour of eigenvalue combinations with $b=-2$. This bias can be represented by an intrinsic probability distribution $p_{\mathrm{int}}(m)$ with $p_{\mathrm{int}}(m)=p_{\mathrm{int}}(\mbox{high})$ for $b=-2$ and $p_{\mathrm{int}}(m)=p_{\mathrm{int}}(\mbox{low})$ for $b=+2$. We can use this fully defined intrinsic probability distribution to explain the effects of errors in terms of a "bit-flip model" where the visibilities are explained by random flips of the intrinsic polarization values \cite{suzuki2012violation}. In the present situation, the probability of each outcome $p(m)$ depends only on the probability $p_{\mathrm{bflip}}(m)$ that the combination of bit flips in $x_A$, $y_A$, $x_B$, and $y_B$ result in a change of the value of $b$. If $m_{+2}$ represents outcomes with $b=+2$ and  $m_{-2}$ represents outcomes with $b=-2$, the outcome dependent probabilities of changes in the $b$-value determine the experimentally observed probability as  
\begin{eqnarray}
p(m_{+2}) &=& (1 - p_{\mathrm{bflip}}(m_{+2})) \cdot p_{\mathrm{int}}(\mbox{low}) + p_{\mathrm{bflip}}(m_{+2}) \cdot p_{\mathrm{int}}(\mbox{high}) \nonumber \\
p(m_{-2}) &=& (1 - p_{\mathrm{bflip}}(m_{-2})) \cdot p_{\mathrm{int}}(\mbox{high}) + p_{\mathrm{bflip}}(m_{-2}) \cdot p_{\mathrm{int}}(\mbox{low}). 
\label{eqn:intrinsic and experimental prob of b=+2}
\end{eqnarray}
Although it is possible to identify the theoretically expected relation between bit flip errors in $x_A$, $y_A$, $x_B$, and $y_B$ with the error probability $p_{\mathrm{bflip}}(m)$ that these errors change the value of $b$, it is sufficient for the present analysis to understand that the changes in outcome probabilities $p(m_{\pm 2})$ caused by changes in measurement uncertainties originate only from the errors in the value of $b$ determined from the joint measurement results. 
If all visibilities are equal, the experimental results show that the error probability $p_{\mathrm{bflip}}(m)$ is independent of the outcome $m$ and can be derived from the ratio of average $\langle b \rangle$ and expectation value $\langle \hat{B} \rangle$,
\begin{equation}
p_{\mathrm{bflip}} = \frac{1}{2}\left(1-\frac{\langle b \rangle}{\langle \hat{B} \rangle}\right).
\label{eqn: measurement visibility of b}
\end{equation} 
The theoretically expected value of $p_{\mathrm{bflip}}$ for equal visibilities is $0.25$, corresponding to $\langle b \rangle=0.5 \langle \hat{B} \rangle$. Assuming a maximal Bell's inequality violation, the present experiment achieves $\langle b \rangle=(0.487 \pm 0.001) \langle \hat{B} \rangle$. The experimental value of $p_{\mathrm{bflip}}$ is then equal to $0.2565 \pm 0.0005$.  
Eq.(\ref{eqn: measurement visibility of b}) assumes that the violation of Bell's inequality corresponds to a negative intrinsic probability of the outcomes with $b=+2$. Specifically, we find that 
\begin{eqnarray}
p_{\mathrm{int}}(\mbox{high}) &=& \frac{1}{16}\left(1 + \frac{|\langle \hat{B} \rangle|}{2}\right)
\nonumber \\
p_{\mathrm{int}}(\mbox{low}) &=& \frac{1}{16}\left(1 - \frac{|\langle \hat{B} \rangle|}{2}\right).
\label{eqn:pint}
\end{eqnarray}
For Bell's inequality violations, the error probabilities $p_{\mathrm{bflip}}$ must be sufficiently high to avoid negative measurement probabilities in Eq.(\ref{eqn:intrinsic and experimental prob of b=+2}). The upper limit of Bell's inequality violations for a given value of $p_{\mathrm{bflip}}$ is
\begin{equation}
p_{\mathrm{bflip}}(m_{+2}) > \frac{|\langle \hat{B} \rangle| - 2}{2 |\langle \hat{B} \rangle|}.
\label{eqn:Climit}
\end{equation}
It is therefore possible to identify an upper limit of the Bell's inequality violation by finding the minimal error probability $p_{\mathrm{bflip}}(m_{+2})$ for a specific low probability measurement result $m_{+2}$. In the following, we investigate the effects of the uncertainty trade-off on the error probabilities $p_{\mathrm{bflip}}$, as evidenced by the changes in count rates observed for the different measurement outcomes.

\subsection{Combination of high $\hat{X}$ resolution with non-vanishing $\hat{Y}$ resolution} 
\label{subsec:distributions of low x}

At $\theta = 45^\circ$, error probabilities are evenly distributed over all outcomes. The extreme opposite is observed at $\theta  = 0^\circ$, where there are no errors in $\hat{X}$ and the measurement outcomes $y$ are completely random. The transition from $\theta = 0^\circ$ to $\theta = 45^\circ$ is characterized by $V_X > V_Y$, with rapidly increasing values of $V_Y$ and gradually decreasing values of $V_X$. 

\begin{figure}[h]
	\centering
	\includegraphics[width=100mm]{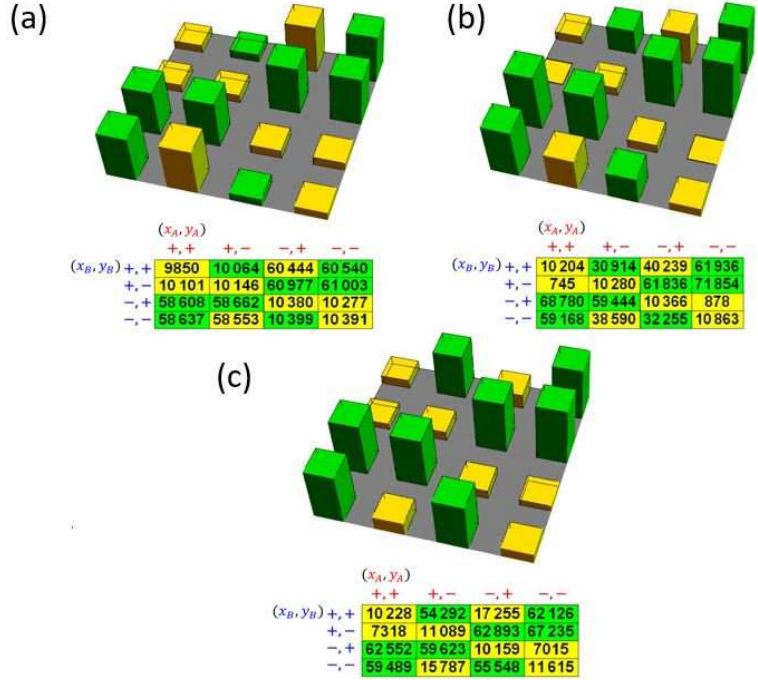}
	\caption{Raw data of the joint statistical distributions measured for uncertainty trade-offs of (a) $\theta = 0^\circ$, $V_X=1$, $V_Y=0$, (b) $\theta = 20^\circ$, $V_X=0.94$, $V_Y=0.34$ and (c) $\theta = 40^\circ$, $V_X=0.77$, $V_Y=0.64$. Outcomes with $b=+2$ are shown in yellow and outcomes with $b=-2$ are shown in green. The numbers given in the table are coincidence counts detected in 10 seconds. Outcomes are labeled in red for $A$ and in blue for $B$, with $+$ and $-$ used in place of the values $+1$ or $-1$. The bar graph is added to better visualize the distribution given in the table.} 
	\label{fig:statistical distributions low x}
\end{figure}

Fig. \ref{fig:statistical distributions low x} shows the experimental count rates obtained at $\theta  = 0^\circ$, $\theta  = 20^\circ$ and $\theta  = 40^\circ$. At $\theta  = 0^\circ$, high count rates are observed for $x_A x_B=-1$ and low count rates are observed for $ x_A x_B=+1$. Among the results with $x_A x_B=-1$, six results have $b=-2$ and two have $b=+2$. The two results with $b=+2$ rapidly decrease in probability as the visibility $V_Y$ increases. Likewise, the low probability of the two outcomes with $x_A x_B=+1$ and $b=-2$ rapidly increases as $V_Y$ increases. 

Less obvious but perhaps more important is the behavior of low probabilities of the outcomes with $y_A y_B=-1$ and $b=+2$. The experimental results show that these probabilities drop from 1010.1 cps at $\theta  = 0^\circ$ to 74.5 cps at $\theta  = 20^\circ$ for $(+,+;+,-)$ and from 1027.7 cps at $\theta  = 0^\circ$ to 87.8 cps at $\theta  = 20^\circ$ for $(-,-;-,+)$. The count rates then return to higher values of 731.8 cps and 701.5 cps at $\theta  = 40^\circ$. The experimental evidence thus indicates that these results achieve much lower values of $p_{\mathrm{bflip}}(m_{+2})$ than the results at $\theta  = 45^\circ$. 

The bit flip model can explain the origin of this low value of $p_{\mathrm{bflip}}(m_{+2})$. At $\theta = 20^\circ$, the visibility of $V_X=0.94$ indicates an error probability of only 0.03 in each $\hat{X}$. On the other hand, the visibility of $V_Y=0.34$ means that simultaneous errors in both $\hat{Y}_A$ and $\hat{Y}_B$ occur with a probability of only 0.0256. Most of the errors originate from individual bit flips of the outcomes $y_A$ and $y_B$. However, such errors do not change the $b$-value of the outcome. The increase in visibility $V_Y$ therefore reduces the probability of errors $p_{\mathrm{bflip}}(m_{+2})$, achieving a minimum near $\theta = 20^\circ$. The effects of errors in $\hat{X}$ ensure that the error probability rises again as $V_X$ begins to decrease, reaching a value of $V_X=0.77$ at $\theta = 40^\circ$.  

Probabilities can be obtained from the count rates as relative frequencies. The total count rate of 568352 counts in 10 seconds can be obtained by adding all of the counts observed for the 16 outcomes. The probabilities of the two outcomes with the lowest count rates are then given by $p(+,+;+,-)=0.001311$ and $p(-,-;-,+)=0.001545$, respectively. The closeness of these probabilities to zero should be evaluated in terms of the unbiased probability of 1/16 for an equal distribution of all 16 possible outcomes. These ratios are given by $16 p(+,+;+,-)=0.02098$ and $16 p(-,-;-,+)=0.02472$, representing a background error of two to three percent that is consistent with the two photon visibilities of the entangled photon source.

\subsection{Combination of non-vanishing $\hat{X}$ resolution with high $\hat{Y}$ resolution}
\label{subsec:distributions of low y}

\begin{figure}[h]
	\centering
	\includegraphics[width=100mm]{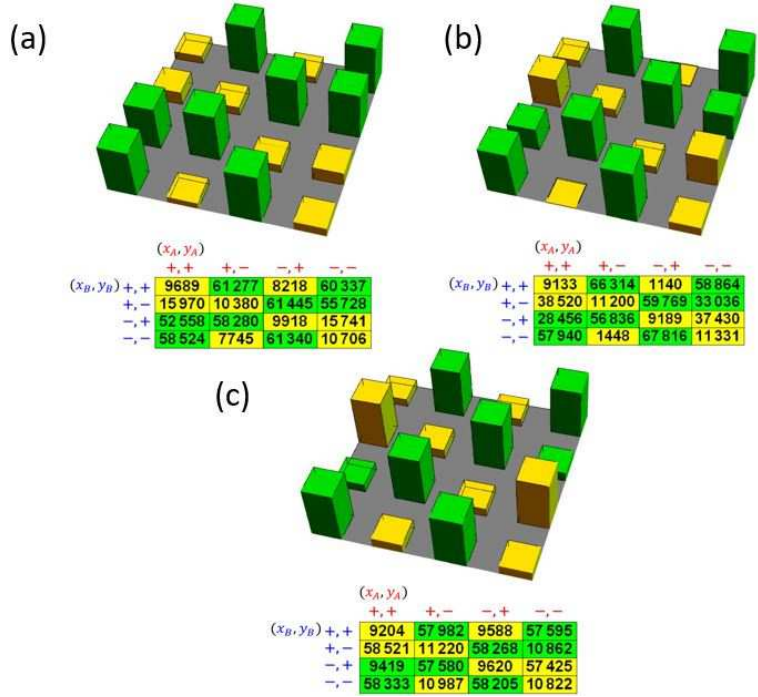}
	\caption{Raw data of the joint statistical distributions measured for uncertainty trade-offs of (a) $\theta = 50^\circ$, $V_X=0.64$, $V_Y=0.77$, (b) $\theta = 70^\circ$, $V_X=0.34$, $V_Y=0.94$ and (c) $\theta = 90^\circ$, $V_X=0$, $V_Y=1$. Outcomes with $b=+2$ are shown in yellow and outcomes with $b=-2$ are shown in green. The numbers given in the table are coincidence counts detected in 10 seconds. Outcomes are labeled in red for $A$ and in blue for $B$, with $+$ and $-$ used in place of the values $+1$ or $-1$. The bar graph is added to better visualize the distribution given in the table.} 
	\label{fig:statistical distributions low y}
\end{figure}

For the sake of completeness it is interesting to consider the results observed at $\theta > 45^\circ$, where $V_X < V_Y$. Fig. \ref{fig:statistical distributions low y} shows the experimental count rates obtained at $\theta  = 50^\circ$, $\theta  = 70^\circ$ and $\theta  = 90^\circ$. At $\theta  = 90^\circ$, high count rates are observed for $y_A y_B=-1$ and low count rates are observed for $ y_A y_B=+1$. It is interesting to note that the results with $y_A y_B = -1$ and $b=+2$ that resulted in minimal experimental probabilities at $\theta = 20^\circ$ increase rapidly between $\theta = 50^\circ$ and $\theta = 90^\circ$, achieving high outcome probabilities at $V_X=0$ and $V_Y=1$. Likewise the probabilities for $x_A x_B = -1$ and $b=+2$ that dropped rapidly between $\theta = 0^\circ$ and $\theta = 40^\circ$ now achieve minimal count rates at $\theta = 70^\circ$, where $V_X = 0.34$ and $V_Y = 0.94$. As discussed in subsec. \ref{subsec:distributions of low x}, this behavior can be explained by the suppression of double errors in both $x_A$ and $x_B$. 

\vspace{\baselineskip}

There is a direct relation between the rapid change of probability in one sector of the uncertainty trade-off and the achievement of minimal probabilities in the other sector. For $b=+2$, low count rates are observed at $\theta = 45^\circ$. For $x_A x_B = -1$, high count rates are observed at $\theta = 0^\circ$ and low count rates are observed at $\theta = 90^\circ$. The count rates therefore drop continuously from $\theta = 0^\circ$ to $\theta = 45^\circ$, have a minimum near zero at $\theta = 67.5^\circ$, and rise back up to a low but non-vanishing value at $\theta = 90^\circ$. A similar behavior is observed for $y_A y_B = -1$, with high count rates at $\theta = 90^\circ$ and low count rates at $\theta = 0^\circ$. Count rates drop to a minimum close to zero at $\theta = 22.5^\circ$ and then rise continuously from $\theta = 45^\circ$ to $\theta = 90^\circ$. 

The probabilities of the outcomes with low count rates can be determined using the total count rate of 548422 counts in 10 seconds obtained by adding the counts of the 16 outcomes. The probabilities associated with the lowest count rates are $p(+,+;-+)=0.002079$ and $p(-,-;+,-)=0.002640$, respectively. The closeness of these probabilities to zero should be evaluated in terms of the unbiased probability of 1/16 for an equal distribution of all 16 possible outcomes. These ratios are given by $16 p(+,+;+,-)=0.03326$ and $16 p(-,-;-,+)=0.04224$. The background errors are a bit higher than the ones observed at $\theta = 20^\circ$, but the magnitude is still consistent with the observed visibilities.

\subsection{Relation between Cirel'son bound and measurement uncertainty limit}

Eq.(\ref{eqn:Climit}) shows that a Bell inequality violation of $\langle \hat{B} \rangle > 2$ requires a minimal error probability $p_{\mathrm{bflip}}(m_{+2})$ for all low probability outcomes $m_{+2}$. This requirement can only be satisfied because the uncertainty principle prevents a suppression of $p_{\mathrm{bflip}}(m_{+2})$ to zero. The general theory of this relation between measurement uncertainties and the Cirel'son bound has been discussed in terms of the visibilities of joint measurements in \cite{hofmann2019local}. The present experimental results show that the effects of the visibilities with regard to the Cirel'son bound can be summarized using the probabilities $p_{\mathrm{bflip}}(m_{+2})$ of obtaining an incorrect value of $b$ due to a combination of bit flip errors. It is possible to derive the precise error probability $p_{\mathrm{bflip}}(m_{+2})$ from the visibilities for every measurement outcome $m_{+2}$. In the present experiment, minimal error probabilities were observed for the outcomes $m_{+2}=(+,+;+,-)$, $m_{+2}=(-,-;-,+)$, $m_{+2}=(-,+;+,+)$, and $m_{+2}=(+,-;-,-)$. For these four outcomes, the error probabilities are given by 
\begin{eqnarray}
  \fl
  p_{\mathrm{bflip}}(+,+;+,-) &=& p_{\mathrm{bflip}}(-,-;-,+) = \frac{1}{4}(2 - V_X^2-2 V_X V_Y + V_Y^2) 
  \label{eqn:error prob of pppm and mmmp} \\
  \fl
  p_{\mathrm{bflip}}(-,+;+,+) &=& p_{\mathrm{bflip}}(+,-;-,-) = \frac{1}{4}(2 + V_X^2 -2 V_X V_Y - V_Y^2).
  \label{eqn:error prob of mppp and pmmm}
\end{eqnarray}
It is a straightforward matter to confirm that the minimal value of these error probabilities is $(2-\sqrt{2})/4$ or about 0.1464. For the settings of $\theta=20^\circ$ and $\theta=70^\circ$ used in the present experiment, the visibilities given by Eq.(\ref{eqn:theta}) give a value of $p_{\mathrm{bflip}}(m_{+2})=0.1478$. This result is only slightly higher than the actual minimum at $\theta=22.5^\circ$, indicating that the observation of the Cirel'son bound is robust against small changes in the measurement settings described by $\theta$. 
 
\begin{figure}[h]
	\centering
	\includegraphics[width=100mm]{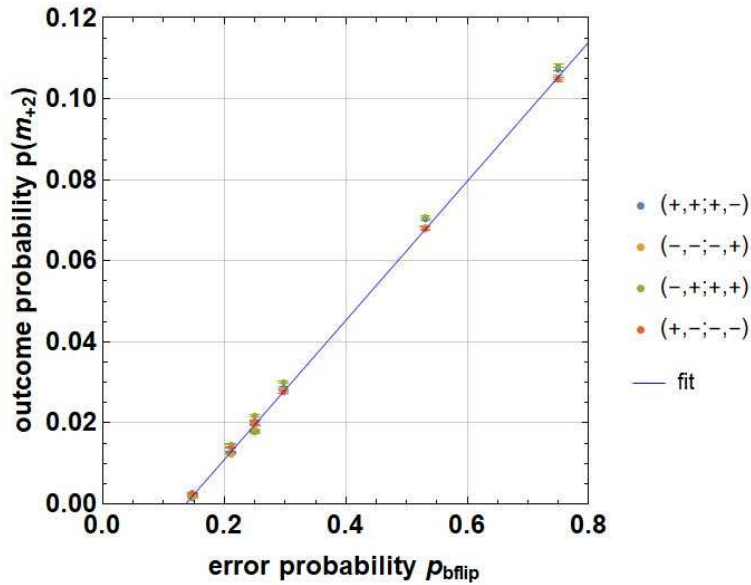}
	\caption{Experimental joint probabilities of the four outcomes $m_{+2}$ that achieve minimal probabilities in the joint measurements plotted as a function of the error probabilities $p_{\mathrm{bflip}}(m_{+2})$ associated with each measurement setting $\theta$. The error probability $p_{\mathrm{bflip}}(m_{+2})$ explains the observed probabilities in terms of their relation with the intrinsic statistics given by the expectation value $|\langle \hat{B} \rangle|$. The blue line shows the best linear fit of the data with a slope of $|\langle \hat{B} \rangle|/16=0.1717$.}
	\label{fig:pbe versus joint prob}
\end{figure}

The Cirel'son bound is determined by the linear relation between the experimentally observed probabilities $p(m_{+2})$ and the error probabilities $p_{\mathrm{bflip}}(m_{+2})$ of the same outcome. This relation is given by Eq.(\ref{eqn:intrinsic and experimental prob of b=+2}), where the linear coefficients are given by the intrinsic probabilities $p_{\mathrm{int}}(\mbox{high})$ and $p_{\mathrm{int}}(\mbox{low})$. These probabilities are directly determined by the expectation value $\langle \hat{B} \rangle$ as shown in Eq.(\ref{eqn:pint}). We can now express the relation between the Cirel'son bound and the error probabilities in $b$ directly in terms of the experimentally observable linear relation
\begin{equation}
p(m_{+2})=\frac{1}{16}\left(|\langle \hat{B} \rangle| p_{\mathrm{bflip}}(m_{+2}) - \frac{|\langle \hat{B} \rangle| - 2}{2} \right).
\label{eqn:linear}
\end{equation}
We have confirmed this linear relation by determining $p_{\mathrm{bflip}}(m_{+2})$ for the various settings of $\theta$ using the visibilities obtained at these settings and relating them to the experimentally observed probabilities of the corresponding outcomes. The result is shown in Fig. \ref{fig:pbe versus joint prob}. 

\vspace{\baselineskip}

It is possible to determine the expectation value $|\langle \hat{B} \rangle|$ from the slope of the relation between $p(m_{+2})$ and $p_{\mathrm{bflip}}(m_{+2})$ shown in Fig. \ref{fig:pbe versus joint prob}. The slope obtained from the experimental data is
$|\langle \hat{B} \rangle|/16 = 0.17173 \pm 0.00017$, corresponding to an expectation value of $|\langle \hat{B} \rangle| = 2.7476 \pm 0.0027$. The relative closeness of this value to the Cirel'son bound is given by 
\begin{equation}
  \frac{|\langle \hat{B} \rangle|}{2 \sqrt{2}} = 0.9716 \pm 0.0011. 
  \label{eqn: ratio between Cirel'son bound and exp average}
\end{equation}
Note that this value is obtained as an average over the joint measurement results obtained at different measurement uncertainty trade-offs $\theta$, not just the ones for which the outcome probabilities are minimal. The result is consistent with the two photon visibilities observed when we evaluated the entanglement of our source. Specifically, the ratio between the experimentally observed expectation value $|\langle \hat{B} \rangle|$ and the pure state limit of $2 \sqrt{2}$ is only about $0.4 \%$ lower than the average of the four interferometer visibilities given in Eq.(\ref{eqn:inter visi}), which is equal to $0.9750 \pm 0.0020$. 

Finally it might be of interest to consider the extrapolation of the linear relation in Eq.(\ref{eqn:linear}) to error probabilities of zero. This extrapolation results in negative probabilities as a direct consequence of the violation of Bell's inequalities. As shown by Eq.(\ref{eqn:intrinsic and experimental prob of b=+2}), the value obtained is equal to the intrinsic probability $p_{\mathrm{int}}(\mbox{low})$. For the present data, 
\begin{equation}
p_{\mathrm{int}}(\mbox{low}) = - 0.02336 \pm 0.00008.
\end{equation}
The violation of Bell's inequality therefore emerges naturally when measurement uncertainties are mathematically removed from the statistics of joint measurements. In the conventional verifications of Bell's inequality violations, the probabilities of measurement outcomes are positive because each measurement is only sensitive to $\hat{X}$ or to $\hat{Y}$. In joint measurements, this is ensured by the surprisingly complicated interplay between the different measurement uncertainties summarized here as a minimal value of the error probability $p_{\mathrm{bflip}}(m_{+2})$.

\section{Conclusions}
\label{sec:conclusion}

We have investigated the relation between the measurement uncertainties of joint measurements and the non-local correlations responsible for the violation of Bell's inequalities by experimentally observing the joint probabilities of the four outcomes $x_A$, $y_A$, $x_B$, and $y_B$ that appear in the formulation of Bell's inequalities for an input state that maximally violates the corresponding Bell's inequality. By varying the uncertainty trade-off in our measurements we have identified the maximal sensitivity of experimentally observed outcome probabilities to the total correlation $|\langle \hat{B} \rangle|$ that appears in Bell's inequalities. 

The experimental result obtained at equal uncertainties in $\hat{X}$ and $\hat{Y}$ shows that the probabilities of the outcomes are directly related to the classical calculation of a $b$-value of $+2$ or $-2$ from the four outcome values determined in the joint measurement. This result strongly suggests that Bell's inequalities can be explored by using a statistical error model to represent the measurement uncertainties, where the statistics of the input state are represented by non-positive intrinsic probabilities \cite{suzuki2012violation,suzuki2016observation}. We can therefore conjecture that the violation of Bell's inequalities will be limited by the possibility of observing the non-local correlations in joint measurements, where the upper limit of the Bell's inequality violation should correspond to the observation of a joint probability of zero for some of the measurement outcomes. 

We have identified four measurement outcomes that achieve probabilities close to zero in the joint measurements. Two of these outcomes have minimal probabilities when the measurement uncertainty of $\hat{X}$ is very low and the other two have minimal probabilities when the uncertainty of $\hat{Y}$ is very low. The reason for this lack of symmetry between $\hat{X}$ and $\hat{Y}$ is the different probability $p_{\mathrm{bflip}}(m_{+2})$ of observing the wrong $b$-value for the different low probability outcomes $m_{+2}$. Measurement uncertainties define a non-zero lower bound for the error probability $p_{\mathrm{bflip}}(m_{+2})$, but it is significantly lower than the homogenous error probability of 0.25 obtained when the measurement uncertainties of $\hat{X}$ and $\hat{Y}$ are all equal. 

We have experimentally confirmed that the probabilities obtained in the joint measurements can be explained by the error probabilities of a basic bit flip model. The extrapolation to error probabilities of zero would then result in unphysical negative probabilities, demonstrating the need for minimal measurement uncertainties in joint measurements of $\hat{X}$ and $\hat{Y}$. The violation of Bell's inequalities can be confirmed by this extrapolation of the statistics, with a value of $|\langle \hat{B} \rangle| = 2.7476 \pm 0.0027$ obtained in the present measurement. 

In conclusion, the statistics of uncertainty limited joint measurements reveals that the statistics of $\hat{B}$ can be explained in terms of the individual values $b$ of joint measurement outcomes, where measurement uncertainties are necessary to avoid any direct observation of the negative intrinsic probabilities required to explain the violation of Bell's inequalities. As the analysis of the experimental results has shown, the Cirel'son bound is imposed by the possibility of observing joint statistics at an optimized uncertainty trade-off within the limits set by the uncertainty relations of the joint measurement. The uncertainty principle of quantum measurements makes it impossible to identify measurement independent realities, explaining why it is so difficult to apply common sense notions of reality to quantum systems. Its application to the failure of local realism represented by the violation of Bell's inequalities shows that the relation between experimental results and objective reality needs to be reformulated in a manner that can accommodate the wide range of possible measurements that can be applied to quantum systems.

\section*{References}
\bibliography{references}

\end{document}